\documentclass[12pt,a4paper]{article}

\usepackage{jheppub}

\usepackage{amsmath,mathtools,color} 
\usepackage{amsfonts} 
\usepackage{amssymb}
\usepackage{graphicx}
\usepackage{caption}
\usepackage{subcaption}

\usepackage{microtype}
\usepackage{color} 

\newcommand{\pt}{\partial}
\newcommand{\M}{M_{\textrm{Pl}}}
\newcommand{\K}{\mathcal K}
\newcommand{\U}{\mathcal U}
\newcommand{\I}{\mathcal I}
\newcommand{\<}{\left\langle}

\DeclareMathOperator{\tr}{tr}

\renewcommand{\[}{\left[}
\renewcommand{\]}{\right]}
\renewcommand{\L}{\mathcal L}
\renewcommand{\>}{\right\rangle}
\renewcommand{\O}{\mathcal O}

\newcommand{\be}{\begin{equation}}
\newcommand{\ee}{\end{equation}}

\title{Stability of Massive Gravity Solutions for Holographic Conductivity}
\author{Lasma Alberte, }
\author{Andrei Khmelnitsky}
\affiliation{Department of Physics, Ben-Gurion University, Beer Sheva, 84105, Israel}
\emailAdd{lasma@post.bgu.ac.il}
\emailAdd{andreykh@post.bgu.ac.il}

\abstract{
We consider non-linear massive gravity with two St\"uckelberg fields in which the diffeomorphism invariance is broken spontaneously in two of the spacetime directions. This theory admits a charged anti-de Sitter black brane solution and has recently been used in holographic context as a bulk description of a boundary field theory with momentum dissipation. Here we study the stability of the black brane solution. We identify the physical degrees of freedom and determine the healthy regions of the parameter space of the theory. We find that there is a region in parameter space, where the theory suffers from ghost instability. We recognise this as the reason for the instability found in the previous works on holographic conductivity. We also rederive the previous results for the holographic conductivity in a coordinate independent way. 
}

\begin{document}
\maketitle
\flushbottom

\section{Introduction}
In the past years there has been a lot of interest in the theory of massive gravity. Phenomenologically, the recent development has been motivated by the expectations that a graviton with a small mass might provide a dynamical explanation of the late accelerated expansion of our Universe. On the other hand, the revived interest on the theoretical side is due to the considerable progress that has been made in working out a theory that can consistently describe a massive spin-2 field. As of today, there is one class of Lorentz invariant massive gravity models proposed by de Rham, Gabadadze, and Tolley (dRGT) that are believed to be healthy at least on the Minkowski background~\cite{deRham:2010kj} (for a recent review, see~\cite{deRham:2014zqa}). 

The theory of massive gravity necessarily invokes a reference metric, usually denoted by $f_{\mu\nu}$~\cite{Boulware:1973my}. In a diffeomorphism invariant theory the reference metric can be treated in two ways. The first option is to attribute it to another dynamical spin-2 tensor field and treat it in the context of a bimetric theory. This is a whole separate, although closely related, field of study with the first ghost-free Lagrangian proposed in~\cite{Hassan:2011zd}. The other possibility is to restore the gauge invariance of the reference metric by introducing four St\"uckelberg scalar fields corresponding to the Goldstone bosons of the four broken coordinate transformations as~\cite{ArkaniHamed:2002sp,Dubovsky:2004sg,Chamseddine:2010ub}:
\be
f_{\mu\nu} = \pt_\mu\phi^A \pt_\nu\phi^B f_{AB}(\phi) \; .
\ee
In the latter case one retains a configuration space reference metric $f_{AB}(\phi)$, which sets a global symmetry of the theory given by its isometry group~\cite{Alberte:2011ah}. In principle, the freedom in choosing the metric $f_{AB}$ is unlimited. Since the global symmetry sets the isometries of the vacuum of the theory, then in most of the cases one chooses the reference metric to be the Minkowski metric leading to a global Poincar\'e symmetry. For cosmological applications the reference metric is often set to a FRW, in particular, de Sitter metric~\cite{Alberte:2011ah, deRham:2012kf,Langlois:2012hk,Sasaki:2013nka}. 

Another possibility is to change the dimension of the reference metric together with the number of the St\"uckelberg scalars. Models with massive gravity and a number of fields exceeding the number of space-time dimensions was considered in~\cite{Gabadadze:2012tr,Andrews:2013ora}. A complementary option is to reduce the number of St\"uckelberg fields to be less than the dimensionality of the space-time~\cite{Alberte:2013sma}. Considering the Lorentz invariance of the conventional $3+1$ dimensional massive gravity a natural choice among the possible reference metrics with reduced dimensionality is a $3$-dimensional Euclidean metric $f_{AB}=\textrm{diag}(1,1,1)$. This breaks the global symmetry group down to the $SO(3)$ group of spatial rotations and, around the vacuum, this gives rise to the so called Lorentz violating (LV) theories of massive gravity first discussed in~\cite{Rubakov:2004eb,Dubovsky:2004sg} (for a review, see~\cite{Rubakov:2008nh}; for a recent work on the high energy completion of the LV theories see~\cite{Blas:2014ira}). Due to the absence of the $\phi^0$ field in this case, the reduced massive gravity reproduces only a subclass of the LV theories of~\cite{Rubakov:2004eb} with $m_0^2=m_1^2=m_4^2=0$. 

In this work we shall consider the case of a reduced massive gravity with two St\"uckelberg fields previously studied in~\cite{Alberte:2013sma}. Here we extend the theory with a negative cosmological constant and a minimally coupled $U(1)$ gauge field.\footnote{Charged black hole solutions in the full dRGT theory have been previously studied in~\cite{Cai:2012db}.}
This theory can be used in the context of the AdS/CFT correspondence as a gravity dual of a field theory with broken translational invariance~\cite{Vegh:2013sk}. The resulting momentum dissipation on the field theory side is needed in order to describe materials with finite conductivity. In a field theory with translational invariance the momentum is conserved. Therefore, in the presence of finite charge density the system has divergent conductivity at zero frequency due to the absence of the possibility to dissipate the current. In order to incorporate the effects of momentum dissipation in the holographic description, it is useful to notice that the conservation of the energy-momentum tensor in the boundary theory arises due to the diffeomorphism invariance of the gravitational theory in the bulk. Since the diffeomorphism invariance is broken in the unitary gauge of the massive gravity, it is a natural candidate for a bulk description of a boundary theory in which the momentum is not conserved. This idea was proposed and successfully implemented by Vegh in~\cite{Vegh:2013sk} where he demonstrates numerically that the optical conductivity in massive holography is indeed finite and exhibits a peak at zero frequency. Further analytic studies were done in~\cite{Davison:2013jba,Blake:2013bqa}. The full set of the thermo-electric transport coefficients was computed in~\cite{Amoretti:2014zha,Amoretti:2014mma}. In order to spontaneously break the diffeomorphism invariance in the bulk one can consider a wider class of theories with scalar fields admitting spatially inhomogeneous vacuum solutions. Some of such theories, which, however, do not include the dRGT mass term, were used in the holographic context in~\cite{Andrade:2013gsa,Taylor:2014tka, Baggioli:2014roa}. In a recent paper~\cite{Blake:2013owa}, it was shown that also the equations describing the perturbations of a holographic lattice coincide with those arising in massive gravity.

In this paper we shall investigate the stability of the field perturbations around the black brane solution used in the calculations of the conductivity in~\cite{Vegh:2013sk}. For this it is useful to first recall what we know about the number of degrees of freedom propagated in massive gravity with two St\"uckelberg fields, without adding the Maxwell field. In~\cite{Alberte:2013sma} we performed the full Hamiltonian analysis of this theory and found that both scalar fields induce additional propagating degrees of freedom. In combination with the two polarisations of the massless graviton this leads to $2+2=4$ propagating degrees of freedom in total. This is different from the dRGT massive gravity, where one naively expects $2+4=6$ propagating degrees of freedom, the sixth one being the Boulware-Deser (BD) ghost. Instead, only three combinations of the four St\"uckelberg scalar fields propagate leading to $2+3=5$ degrees of freedom and the absence of the BD ghost~\cite{Hassan:2011hr}. That this is true in the dRGT theory is the best to see in the ADM formulation in unitary gauge where all the degrees of freedom are encoded in the metric. The principal steps of the proof can be outlined as follows. In General Relativity, only six of the metric components enter the Lagrangian with time derivatives. This fact is not changed by adding the graviton mass term since it contains no derivatives of the metric. Therefore, in the absence of the diffeomorphism invariance, six is the maximal amount of degrees of freedom that can be propagated in any sort of massive gravity. In order to show that there is no BD ghost one has to show that the Hamiltonian constraint is preserved and thus eliminates the sixth degree of freedom~\cite{Hassan:2011hr}. In~\cite{Vegh:2013sk} Vegh has presented a version of this proof for the case of a degenerate reference metric $f_{\mu\nu}=\textrm{diag }(0,0,1,1)$ (this is equivalent to the case of massive gravity with two St\"uckelberg fields in unitary gauge). He finds that one of the six degrees of freedom is eliminated due to the fact that the corresponding diffeomorphism remains unbroken.\footnote{The precise statement is that the shift $N^r$ does not appear in the mass term, see~\cite{Vegh:2013sk} for details. } He also shows that the Hamiltonian constraint is preserved in the same manner as it is preserved in the dRGT theory, thus removing one more degree of freedom. He concludes that there are four propagating degrees of freedom in massive gravity with the degenerate reference metric. In the diffeomorphism invariant formulation of this theory there are only two St\"uckelberg fields, and the absence of the two degrees of freedom is trivial. Hence, the issue of the BD ghost is irrelevant in massive gravity with only two St\"uckelberg fields. In this paper we shall investigate the stability of the degrees of freedom present in the reduced massive gravity together with a Maxwell field. 

In addition, we extend the previous calculation of the conductivity in the holographic framework to the diffeomorphism invariant case. For this we shall perform the full analysis of the field perturbations. We shall do so by decomposing the perturbations according to the irreducible representations of the isometry group. We will derive the quadratic action in terms of the diffeomorphism invariant fields. As a result we find that the theory propagates in total six degrees of freedom. Two of them are the usual two tensor polarisations of the graviton, which have become massive due to the addition of the graviton mass term. As expected, two vector degrees of freedom are propagated by the Maxwell field, and another two vector degrees of freedom are propagated by the St\"uckelberg fields. We find no propagating degrees of freedom in the scalar sector of the perturbations. The stability of the dynamics of these fields depends on the parameters of the theory. In particular, we find that for the range of parameters excluded as unstable in~\cite{Vegh:2013sk,Davison:2013jba,Blake:2013bqa} the vector degrees of freedom induced by the St\"uckelberg fields become ghosts. As a consistency check we show that the resulting gauge invariant equations of motion for the vector perturbations coincide with the equations previously derived in~\cite{Vegh:2013sk,Davison:2013jba,Blake:2013bqa} in the unitary gauge. Finally, we calculate the conductivity in the gauge invariant case and show that the results coincide with the previous calculations. 

The paper is organised as follows. In section \ref{sec:model} we present the reduced massive gravity coupled to a Maxwell field and review the properties of the AdS black brane solution. In section \ref{sec:diff} we classify the perturbations of the fields according to the representations of the isometry group and find the diffeomorphism invariant field variables. We derive the full quadratic action in section \ref{sec:action}. The conductivity calculations are presented in section \ref{sec:conductivity}. Section \ref{sec:conclusions} is devoted to onclusions and discussion. Since the massive gravity in~\cite{Vegh:2013sk} is written in slightly different notations, we also present the comparison between the different notations in appendix. 

\section{The model}\label{sec:model}
We consider massive gravity with a negative cosmological constant and a $U(1)$ gauge field described by the Lagrangian:
\be\label{action}
\L=\frac{1}{2}\sqrt{-g}\left[\M^2\left(R+2\Lambda\right)-\frac{L^2}{4e^2}F_{\mu\nu}F^{\mu\nu}+\M^2 \,\U(g,\phi)\right],
\ee
where $\M^2\equiv (8\pi G)^{-1}$, $\Lambda$ is the cosmological constant, and $F_{\mu\nu}$ is the Maxwell field. $\U(g,\phi)$ is the mass potential for the graviton written as
\be
\U(g,\phi) = \beta_1 \, \U_1 + \beta_2 \, \U_2 \;,
\ee
where 
\begin{align}\label{defy}
&\U_1=[\sqrt{\I}] ,\quad \U_2=[\sqrt{\I}]^2 - [\sqrt{\I}\,^2] \; , \\ \label{defiab}
&\I^{AB}\equiv g^{\mu\nu}\pt_\mu\phi^A\pt_\nu\phi^B,\qquad\I^A_B\equiv\I^{AC} f_{BC} \; .
\end{align}
The Greek indices $\mu=\{t,r,x,y\}$ are space-time indices whereas the Latin indices run in the configuration space of the scalar fields $\phi^A$. We will be working with two St\"uckelberg fields $\phi^A$ with $A=\{x,y\}$ and a reference metric $f_{AB}=\delta_{AB}=\mathrm{diag}(1,1)$. In this case $\I^A_B$ is a $2\times 2$ matrix and its square root is defined so that $\sqrt{\I}^A_C\sqrt{\I}^C_B=\I^A_B$. The square brackets denote the traces of the matrix $\sqrt{\I}^A_B$ so that $[\sqrt{\I}]=\sqrt{\I}^A_A,\,[\sqrt{\I}^2]=\sqrt{\I}^A_B\sqrt{\I}^B_A$. An equivalent description in terms of the matrix $\Omega^\mu_\nu\equiv\sqrt{g^{\mu\alpha}f_{\alpha\nu}}$ was used in~\cite{Vegh:2013sk}. The comparison to this approach is presented in appendix \ref{sec:app1}. 

\subsection{Black brane solution}\label{sec:background}
The equations of motion of the St\"uckelberg scalar fields and the Maxwell field admit the background solutions
\be\label{sol}
\phi^A=x^\mu\delta^A_\mu \; ,	\qquad 	A_t = \mu\left(1-\frac{r}{r_h}\right) \;.
\ee
Variation of the action with respect to the metric gives
\be\label{eom}
R_{\mu\nu}-\frac{1}{2}g_{\mu\nu}R-g_{\mu\nu}\Lambda+\frac{L^2}{2}\left(F_{\mu\alpha}F_{\ \nu}^\alpha+\frac{1}{4}g_{\mu\nu}F_{\alpha\beta}F^{\alpha\beta}\right)+X_{\mu\nu}=0 \; ,
\ee
where we have we set $e=\M^2= 1$, and defined
\be\label{xmunu}
X_{\mu\nu}=\beta_1\left[\mathcal{E}_{\mu\nu}-\frac{1}{2}g_{\mu
\nu}\,\U_1\right]+\beta_2\left[\left(2\,\U_1\mathcal E_{\mu\nu}-\mathcal E^2_{\mu\nu}\right)-\frac{1}{2}g_{\mu\nu}\,\U_2\right]
\ee
with
\begin{align}\label{defe1}
&\mathcal E_{\mu\nu}\equiv\frac{\delta [\sqrt{\I}]}{\delta g^{\mu\nu}}= \frac{1}{4}\sqrt{\I}^C_A\,\mathcal I^{-1}_{BC}\left(\pt_\mu\phi^A\pt_\nu\phi^B+\pt_\nu\phi^A\pt_\mu\phi^B\right) \; , \\ \label{defe2}
&\mathcal E_{\mu\nu}^2\equiv\frac{\delta[\sqrt{\I}\,^2]}{\delta g^{\mu\nu}} =\frac{1}{2}\sqrt{\I}^D_A\sqrt{\I}^C_D\,\mathcal I^{-1}_{BC}\left(\pt_\mu\phi^A\pt_\nu\phi^B+\pt_\nu\phi^A\pt_\mu\phi^B\right)\; ,
\end{align}
and $\I^{-1}_{AB}$ --- the inverse of $\I^{AB}$. After substituting the black brane ansatz for the background metric
\be
\label{solmetric}
ds^2=L^2\left(\frac{dr^2}{f(r)r^2}+\frac{-f(r)dt^2+dx^2+dy^2}{r^2}\right)
\ee
on the solutions \eqref{sol}, we find the following equation for the emblackening factor $f(r)$:
\begin{equation}\label{eom1}
2rf'-6f+2(\beta_1Lr+\beta_2r^2)-\frac{\mu^2r^4}{2r_h^2}+2L^2\Lambda=0 \; .
\end{equation}
With the cosmological constant $\Lambda=3/L^2$  these equations admit a solution~\cite{Vegh:2013sk}:
\be\label{fr}
f(r)=1+\beta_1\frac{L}{2}r+\beta_2r^2-Mr^3+\frac{\mu^2}{4r_h^2}r^4 \; ,
\ee
where $M$ is a constant of integration. It can be determined by demanding that $f(r)$ vanishes on the horizon $r=r_h$ to be
\be
M=\frac{1}{r_h^3}+\frac{\beta_1L}{2r_h^2}+\frac{\beta_2}{r_h}+\frac{\mu^2}{4r_h} \; .
\ee
In the absence of the graviton mass, the solution \eqref{fr} reduces to the AdS-Reissner-Nordst\"om charged black brane solution. If, instead, we keep the graviton mass term, but set the Maxwell field to zero, i.e. $\mu=0$ in \eqref{fr}, then the solution reduces to a black brane solution with zero charge. If both, the graviton mass term and the Maxwell field are set to zero, then the solution reduces to the AdS-Schwarzschild black hole solution.  
  
\subsection{Temperature}
The temperature of this black brane solution can be found from the analytical continuation of the metric \eqref{solmetric} by setting $\tau=it$. In order to avoid the conical singularity of this space at the horizon, one must periodically identify the Euclidean time $\tau$. This corresponds to considering the dual theory at a finite temperature, given by the inverse of the Euclidean time period. A general expression of the temperature in terms of the emblackening factor $f(r)$ can be written as (see, for example,~\cite{Hartnoll:2009sz}):
 \be\label{temp}
 T=\frac{|f'(r_h)|}{4\pi}=\frac{12+4\beta_1Lr_h+4\beta_2r_h^2-\mu^2r_h^2}{16\pi r_h} \; .
 \ee
 Further thermodynamic quantities describing this solution can be derived from the on-shell Euclidean action together with the Gibbons-Hawking boundary term and additional counterterms accounting for the UV divergences. The details of this study can be found in~\cite{Blake:2013bqa} (see also~\cite{Adams:2014vza} for the study of the thermodynamic phase structure of the model). For the purposes of this note we will need only the expression for the temperature.   
  
  \subsection{Holographic energy-momentum tensor}

 In~\cite{Davison:2013jba}, a phenomenological conservation law describing the momentum dissipation in the boundary theory of the holographic massive gravity at high temperatures was proposed:
 \be\label{cons}
 \pt_t\<T^{tt}\>=0,\qquad\pt_t\<T^{ti}\>=-\tau^{-1} \<T^{ti}\> \; ,
 \ee
 where $\<T^{ab}\>$, $a=\{t,x,y\}$ denotes the holographic stress-energy tensor and $\tau$ parametrises the momentum relaxation timescale. 
 
 In the absence of the knowledge of the covariant counterterms due to the graviton mass term \eqref{defy} (given explicitly in \eqref{u1u2}), the exact expression for the holographic stress-energy tensor is not known. The usual statement of holographic renormalisation is that the Ward identities for the stress-energy tensor can be derived from the condition of the diffeomorphism invariance of the renormalised boundary action \cite{Bianchi:2001kw}. Equivalently, the Ward identities follow from the boundary expansion of the equations of motion. A recent study of this question in the context of holographic momentum dissipation in other theories with two scalar fields can be found in \cite{Andrade:2013gsa,Davison:2014lua,Taylor:2014tka}.
  
The reduced massive gravity with two St\"uckelberg fields, given in \eqref{action}, is diffeomorphism invariant. In principle, this allows one to perform the standard holographic renormalisation, find the boundary counterterms, and derive the Ward identities.  The conservation equations \eqref{cons} can also be derived in a simpler manner from the hydrodynamic limit of the equations of motion \eqref{eom}--\eqref{defe2}. We leave this for future work.

  \section{Diffeomorphism invariant perturbation theory}\label{sec:diff}
 We consider perturbations around the background solution for the fields appearing in the Lagrangian \eqref{action}. These include: $(i)$ perturbations of the metric $\delta g_{\mu\nu}$ defined as
    \be\label{mpert}
  g_{\mu\nu}=\hat g_{\mu\nu}+\delta g_{\mu\nu} \; ,\qquad \delta g_{\mu\nu}=\frac{L^2}{r^2}h_{\mu\nu} \; ,
  \ee
  $(ii)$ perturbations of the Maxwell field $a_\mu$ defined as
   \be
 A_\mu=\hat A_\mu+a_\mu \; ,
 \ee
 and $(iii)$ perturbations of the St\"uckelberg scalar fields $\pi^A$, $A=\{x,y\}=i$ defined as
 \be\label{pi}
\phi^A=\hat\phi^A+\delta\phi^A\equiv x^\mu\delta_\mu^A+\pi^A \; .
\ee
In total, the fields $\delta g_{\mu\nu},\, a_\mu, \pi^A\equiv\pi^i$ are parametrised by 16 components --- ten components of the metric, four components of the Maxwell field, and two components of the St\"uckelberg fields. These perturbations can be classified with respect to the irreducible representations of the two dimensional rotations in the $\{x, y\}$-plane as scalar, vector and tensor perturbations in the following way:
  \begin{align}
  &\textrm{scalar:}\qquad h_{tt},\,h_{tr},\,h_{rr},\,h\equiv \frac{h_{ii}}{2},\,a_t,\,a_r \; ;\\
  &\textrm{vector:}\qquad h_{ti},\,h_{ri},\,a_i,\,\pi^i \; ;\\\label{tens}
  &\textrm{tensor:}\qquad \bar h_{ij}\equiv h_{ij}-\frac{h_{kk}}{2}\delta_{ij}, \,\bar h_{ii}=0 \; .
  \end{align}
   Hence, out of the 16 fields characterising the field perturbations, there are six fields that belong to the scalar representation, four fields with two components each that belong to the vector representation, and two fields that belong to the tensor representation.

All of these perturbation fields transform under the space-time diffeomorphisms $x^\mu\mapsto\tilde x^\mu=x^\mu+\xi^\mu$, defined by the four components of the vector $\xi^\mu$. By properly choosing the coordinate transformation $\xi^\mu$ we can reduce the number of fields we are dealing with by four. Hence, we expect 12 physical degrees of freedom that are independent of the coordinate system. In order to find these gauge invariant combinations of the perturbations, let us consider the transformation laws of the fields $\delta g_{\mu\nu},\,a_\mu,\,\pi^A$. Under the space-time diffeomorphisms the metric perturbations transform as perturbations of a rank two tensor:
  \be
  \delta g_{\mu\nu}\quad\mapsto\quad\delta \tilde{g}_{\mu\nu}=\delta g_{\mu\nu}-\hat g_{\mu\nu,\gamma}\xi^\gamma-\hat g_{\mu\gamma}\xi^\gamma_{,\nu}-\hat g_{\nu\gamma}\xi^\gamma_{,\mu} \; .
  \ee
  The perturbations of the Maxwell field transform as perturbations of a vector:
   \be
 \tilde a_\mu=a_\mu-\hat A_{\mu,\gamma}\xi^\gamma-\hat A_\gamma\xi^\gamma_{,\mu} \; ,
 \ee
  and each of the two perturbations of the St\"uckelberg scalar fields transforms as a perturbation of a scalar
  \be 
\pi^A\quad\mapsto \quad\tilde\pi^A=\pi^A-\partial_\mu\hat\phi^A\xi^\mu \; .
\ee 
From the last equation we see that by choosing $\xi^\mu=\pi^A\delta_A^\mu$, the scalar field perturbations vanish, i.e. $\tilde\pi^A=0$. This is the \emph{unitary gauge}, the most widely used gauge in massive gravity. In the case of the dRGT massive gravity with four St\"uckelberg fields, this fixes the coordinate system uniquely. However, in the case of the reduced massive gravity with only two St\"uckelberg fields, setting $\tilde\pi^A=0$ leaves two unspecified coordinate transformations. In this paper, instead of fixing a particular gauge we shall work in terms of gauge invariant combinations of the fields defined below. 

  In what follows we shall only consider perturbations independent on the transverse directions, i.e. $\delta g_{\mu\nu}=\delta g_{\mu\nu}(t,r)$, $a_\mu=a_\mu(t,r),$ $\pi^A=\pi^A(t,r)$. This ansatz partially breaks the group of diffeomorphisms and invariance is preserved only under the coordinate transformations with $\xi^\mu=\xi^\mu(t,r)$.  At quadratic level, the equations of motion of the different representations decouple and the three sectors can be considered separately. 

\subsection{Scalar perturbations}
As was already said before, there are in total six components of fields which transform as scalars under the transverse rotations: $h_{tt},h_{tr},h_{rr},h,a_t,a_r$. Two combinations of them can be set to zero by an appropriate choice of the coordinate transformations $\xi^t,\xi^r$. Hence, there should be only four independent gauge invariant combinations of the scalar perturbations. Three of them can be found to be
\begin{align}\label{sc1}
&\bar h_{r}\equiv h_{rr}+\frac{1}{f}\left(rh'-\frac{rf'}{2f}h\right),\\\label{sc2}
&\bar h_t\equiv \left(\frac{1}{f}X\right)'-\frac{2}{f}\dot Y \; , \\ \label{ainv}
&\bar a\equiv a_t'-\dot a_r+\frac{\hat A_t'}{2f}X+\hat A_{t}'\left(\frac{r}{2}h\right)',
\end{align}
where we have introduced the short hand notations
\be
X=h_{tt}+\left(f+\frac{rf}{2}\right)h \; , \qquad Y=h_{tr}+\frac{r}{2f}\dot h \; .
\ee
In principle, there should be one more independent gauge invariant combination of the metric and Maxwell field perturbations. However, we will see below that due to the $U(1)$ symmetry of the Maxwell theory the components $a_t,a_r$ enter the equations of motion only in the combination $\bar a$. Hence, the set of the diffeomorphism invariant fields presented above is sufficient in order to describe the perturbations of our model \eqref{action}. 

In case one would want to choose some coordinate system in a way that entirely fixes the gauge freedom of $\xi^t$ and $\xi^r$, it is useful to observe that under the diffeomorphisms $h\mapsto \tilde h=h+2\xi^r/r$, $X\mapsto\tilde X=X+2f\dot\xi^t$, and $Y\mapsto\tilde Y=Y+f{\xi^t}'$. Convenient choices for the gauge fixing would therefore be to set $h=0, X=0$ or $h=0,Y=0$. We will, however, not fix the gauge here, but rewrite the equations of motion for the metric perturbations in terms of the gauge invariant fields $\bar h_t,\bar h_r,\,\bar a$.
 
 \subsection{Vector perturbations}\label{sec:vectors}
 In the vector sector there are four fields $h_{ti},h_{ri},a_i,\pi^i$, each with two components, and two coordinate transformations $\xi^i$. We therefore expect, in total, six gauge invariant fields organised in three vectors of two components each. We find immediately that the vector $a_i$ is invariant under the coordinate transformations, and, hence is the first gauge invariant vector.  We then also find that the combinations
 \begin{align}\label{vec}
 &\bar h_i\equiv h_{ti}'-\dot h_{ri} \; ,\\
 &\bar\pi^i_r\equiv(\pi^i)'-h_{ri} \; ,\\
 &\bar\pi^i_t\equiv\dot\pi^i-h_{ti}
 \end{align}
 are all diffeomorphism invariant and are related to each other as
\be
\dot{\bar\pi}^i_r-(\bar\pi^i_t)'=\bar h_i \; .
\ee
In principle, all three fields $\bar\pi^i_r,\,\bar\pi^i_r,\,\bar h_i$ appear in the equations of motion. However, one can always eliminate one of them by using the above relation.  We then have three independent vector fields --- $a_i$ and two out of $\{\bar h_i,\,\bar\pi^i_r,\,\bar\pi^i_t\}$ --- representing in total six independent field components.  
 
 \subsection{Tensor perturbations}\label{sec:einst_tens}
 The tensor perturbations $\bar h_{ij}$ are invariant under the diffeomorphism transformations and correspond to the two helicity-two degrees of freedom of the graviton.

\section{Quadratic action}\label{sec:action}
In this section we shall derive a quadratic action for the gauge invariant perturbations around the black brane solution. In order to expand the Einstein action up to the second order in perturbations, we need the second order expansions for the inverse metric perturbations and the perturbations of the square root of the determinant of the metric. These are known to be
\begin{align}
&\delta g^{\mu\nu}=-\tilde h^{\mu\nu}+\tilde h^{\mu\alpha}\tilde h^\nu_\alpha+\mathcal O(\tilde h^3) \;,\\
&\sqrt{-g}=\sqrt{-\hat g}\left(1+\frac{1}{2}\tilde h-\frac{1}{4}\left(\tilde h^{\mu\nu}\tilde h_{\mu\nu}-\frac{1}{2}\tilde h^2\right)\right)+\mathcal O(\tilde h^3) \;,
\end{align}
where $\tilde h_{\mu\nu}\equiv L^2/r^2 h_{\mu\nu}$, and the indices of $\tilde h_{\mu\nu}$ are raised with the background metric $\hat g_{\mu\nu}$ so that $\tilde h^{\mu\nu}=\hat g^{\mu\alpha}\hat g^{\nu\beta}\tilde h_{\alpha\beta}$. The second order gravitational action is
\begin{align}
\mathcal L_{GR}^{(2)}&= \frac{1}{2} \sqrt{- \hat g} \, \Bigg(\frac{1}{2}\nabla_\mu \tilde h_{\alpha\beta}\nabla^\alpha \tilde h^{\mu\beta}-\frac{1}{4}\nabla_\mu \tilde h_{\alpha\beta}\nabla^\mu \tilde h^{\alpha\beta}+\frac{1}{4}\nabla^\alpha \tilde h\nabla_\alpha \tilde h-\frac{1}{2}\nabla^\beta \tilde h^\alpha_\beta\nabla_\alpha \tilde h\Big.\nonumber\\
+&\Big.\tilde h^{\mu\alpha}\tilde h^\nu_\alpha\hat R_{\mu\nu}-\frac{1}{2}\tilde h\tilde h^{\mu\nu}\hat R_{\mu\nu}-\frac{1}{4}\left(2\Lambda+\hat R\right)\left(\tilde h^{\mu\nu}\tilde h_{\mu\nu}-\frac{1}{2}\tilde h^2\right)\Bigg) \;,
\end{align}
where $\Lambda=3/L^2$, and $\hat R_{\mu\nu}$ and $\hat R$ denote the background solutions for the Ricci tensor and Ricci scalar as given by the Einstein equation \eqref{eom}.

In order to perturb the graviton mass term, we notice that, instead of varying the square root matrix $\sqrt\I ^A_B$ directly, one can rewrite the characteristic polynomials in terms of the eigenvalues $\lambda_1,\,\lambda_2$ of the matrix $\I^A_B$ as
\begin{align}
&\U_1(\sqrt{\I})=\tr\sqrt{\I}=\sqrt{\lambda_1}+\sqrt{\lambda_2} \; ,\\
&\U_2(\sqrt{\I})=2\det\sqrt{\I}=2\sqrt{\lambda_1\lambda_2} \; .
\end{align}
The eigenvalues $\lambda_i$ can be expressed in terms of the trace and the determinant of the matrix $\I^A_B$ as \be
\lambda_{1,2}=\frac{1}{2}\left(\tr\I\pm\sqrt{4\det\I}\right).
\ee
It follows that
\be\label{u1u2}
\U_1(\sqrt{\I})=\sqrt{\tr\I+2\sqrt{\det\I}} \; , \qquad
\U_2(\sqrt{\I})=2\sqrt{\det\I} \; .
\ee
Thus, the variation of the graviton mass term can be written in terms of the perturbations of the trace and the determinant of the matrix $\I ^A_B$ instead of the square root matrix $\sqrt{\mathcal \I}^A_B$. The perturbations of $\I^A_B$ up to the second order read: 
\be
\I^A_B=\hat\I^A_B+\,^{(1)}\delta\I^A_B+\,^{(2)}\delta\I^A_B+\mathcal O(\delta^3) \; ,
\ee
where $\hat\I^A_B=r^2/L^2\delta^A_B$ and
\begin{align}
\label{pertI1}
\,^{(1)}\delta\I^A_B&=\left(\hat g^{\mu\nu}\left(\pt_\mu\pi^A\delta_\nu^C+\delta_\mu^A\pt_\nu\pi^C\right)-\tilde h^{\mu\nu}\delta^A_\mu\delta^C_\nu\right)\delta_{BC} \, , \\
\,^{(2)}\delta\I^A_B&=\left(\hat g^{\mu\nu}\pt_\mu\pi^A\pt_\nu\pi^C-\tilde h^{\mu\nu}\left(\pt_\mu\pi^A\delta_\nu^C+\delta_\mu^A\pt_\nu\pi^C\right)+\tilde h^{\mu\alpha}\tilde h^\nu_\alpha\delta^A_\mu\delta^C_\nu\right)\delta_{BC} \, .
\end{align}
It is then straightforward to find the variations of $\tr\I$ and $\det\I$ and to subsequently vary the characteristic polynomials $\U_1$ and $\U_2$ given in \eqref{u1u2}.

The variation of the Maxwell Lagrangian is effortless and we shall not dwell on it here. Below we present the resulting quadratic action for the scalar, vector, and tensor perturbations.

\subsection{Scalar perturbations}
By using the background equation of motion for the emblackening factor $f(r)$ and after numerous integrations by parts the action describing the dynamics of the four gauge invariant scalar modes $\bar h_r,\,\bar h_t$, and $\bar a$ can be simplified to
\begin{equation}\label{action_sc}
\L=\frac{L^2}{4r^4} \left( \left(3+\beta_1Lr+\beta_2r^2\right)f^2\bar h_r^2+\frac{\mu r^4}{r_h}f\bar h_r\bar a+r^4\bar a^2-2rf^2\bar h_r\bar h_t\right).
\end{equation}
This action does not contain any time derivatives and, therefore, there are no dynamical degrees of freedom in the scalar sector of the theory.  
Variation of this action with respect to the fields $\bar a,\,\bar h_r,\,\bar h_t$ yields the following equations of motion
\begin{align}
&2\left(3+\beta_1Lr+\beta_2r^2\right)f\bar h_r+\frac{\mu r^4}{r_h}\bar a-2rf\bar h_t=0 \; ,\\
&\mu f\bar h_r+2r_h\bar a=0 \; ,\\
&\bar h_r=0 \; .
\end{align}
By resolving these equations we see that all the fields are equal to zero.

\subsection{Tensor perturbations}
The quadratic action for the diffeomorphism invariant transverse tensor perturbations $\bar h_{xx}$ and $\bar h_{xy}$ we find to be
\begin{align}\label{action_t}
\L &= \frac{L^4}{4r^4}\Big(\bar h_{xx}\left[\Box+\frac{\beta_1r}{2L}\right]\bar h_{xx}+\bar h_{xy}\left[\Box+\frac{\beta_1r}{2L}\right]\bar h_{xy}\Big) \; ,
\end{align}
where $\Box$ is the scalar d'Alembert operator in the background metric $\hat g_{\mu\nu}$. Hence, there are two propagating degrees of freedom $\bar h_{xy}$ and $\bar h_{xx}$ that satisfy the Klein -- Gordon equation in the background space-time with mass $m^2_\textrm{g}(r) \equiv -\beta_1 r / (2L)$. These degrees of freedom correspond to the graviton of General Relativity, which due to the potential $\U(g,\phi)$ in the gravitational action, has now acquired an effective mass. It seems, that positive values of $\beta_1$ correspond to negative graviton mass squared and therefore will imply a tachyonic instability. However, the mass $m^2_\textrm{g}(r)$ depends on $r$, and the question of stability of the tensor mode requires a more careful analysis.

It is interesting to note that only the $\beta_1$ term contributes to the graviton mass. Hence, by setting $\beta_1$ to zero, the graviton would remain massless. We expect that, in the case when the fields depend on all space-time coordinates (not only on $t$ and $r$), also the $\beta_2$ term contributes to the graviton mass.

\subsection{Vector perturbations}\label{sec:action_vec}
In this work, the dynamics of the vector perturbations is of our greatest interest since they are the most relevant for the calculations of the conductivity in the holographic framework. We find that the quadratic Lagrangian takes the form
\begin{equation}\label{act_vec}
\L=\frac{L^2}{4r^2f}
\left( m^2(r)(\bar\pi_t^i)^2 - m^2(r)f^2(\bar\pi_r^i)^2 - \frac{2\mu r^2}{r_h}fa_i\bar h_i+f\bar h_i^2-r^2f^2 a_i'^2+r^2\dot{a}_i^2\right) \;,
\end{equation}
where we have defined
\be
m^2(r)\equiv - \frac{\beta_1L}{r} - 2\beta_2 \;,
\ee
and summation over the index $i$ is implied. We notice, that all four gauge invariant vector fields $\bar\pi^i_t,\,\bar\pi^i_r,\,\bar h_i,\,a_i$ introduced in the section \ref{sec:vectors} appear in the action. The three fields $\bar\pi^i_r$, $\bar\pi^i_t$, and $\bar h_i$ are, however, related to each other via constraint
\begin{equation}\label{vec_constraint}
\bar h_i=\dot{\bar\pi}_r^i - (\bar\pi_t^i)' \;.
\end{equation}
The vector indices of the fields are always contracted trivially, and in what follows we shall suppress them in order to simplify the appearance of the expressions.

Since one expects at most two dynamical vector fields in the theory -- the Maxwell field and the helicity-one part of the massive graviton -- it should be possible to integrate out two of the four vector fields. In the case at hand, one can express $\bar h$ in terms of $\bar\pi_t$ and $\bar\pi_r$ in the action \eqref{act_vec} and then vary it with respect to $a$, $\bar\pi_t$, and $\bar\pi_r$. The resulting equations of motion do not allow to express any of the fields in terms of the other two algebraically, without the need to invert differential operators. However, it is possible to obtain an action for the dynamical degrees of freedom only. For this one should note, that the fields $\bar\pi_t,\,\bar\pi_r$, and $\bar h$ of the gravity sector appear in the action ~\eqref{act_vec} without derivatives, and their dynamics is only due to the constraint~\eqref{vec_constraint}. One can instead treat these fields as independent and enforce the constraint by adding it to the Lagrangian with a Lagrange multiplier $\lambda$:
\begin{equation}
\L_\lambda = \lambda \left( \bar h - \dot{\bar\pi}_r + (\bar\pi_t)'  \right) \;.
\end{equation}
The action \eqref{act_vec} together with the above constraint contains five vector fields $\bar\pi_t,\,\bar\pi_r,\,\bar h,\,a$, and $\lambda$, all of which should be treated independently. The corresponding equations of motion read
\begin{align}
\bar\pi_r &= \frac{2 r^2}{m^2(r)f} \, \dot\lambda \; ,  \label{vec_eom_r}\\
\bar\pi_t &= \frac{2 r^2 f}{m^2(r)} \, \lambda' \; ,  \label{vec_eom_t}\\
\bar h &= \frac{\mu r^2} {r_h} \, a- 2 r^2 \lambda \; , \label{vec_eom_h}\\
\bar h &= \dot{\bar\pi}_r - (\bar\pi_t)' \; , \\
\ddot{a}_i &- f^2 a_i''-ff'a_i'+\frac{\mu f}{r_h}\bar h_i = 0 \;. 
\end{align}
The first three equations allow to solve for the fields $\bar\pi_t,\,\bar\pi_r$, and $\bar h$ algebraically. By plugging the solutions back in the Lagrangian~\eqref{act_vec} (with the constraint term added) we obtain a Lagrangian for the two dynamical vector degrees of freedom:
\begin{equation}\label{act_lam_a}
\L = L^2 \left ( \frac1{4 f} \, \dot a^2 - \frac f4 \, (a')^2 
+ \frac{r^2}{m^2(r) f} \, \dot \lambda^2 - \frac{r^2 f}{m^2(r)} \, (\lambda')^2
- r^2 \left( \lambda - \frac\mu{2 r_h} \, a \right)^2
\right) \;.
\end{equation}
The two vector degrees of freedom in this Lagrangian are mixed only through the ``mass term''. However, the derivative terms come with different $r$-dependent factors and it is impossible to diagonalise all three quadratic forms corresponding to time derivatives, spatial derivatives, and masses simultaneously. The Lagrangian~\eqref{act_lam_a} also shows that only one linear combination of the fields, namely the one corresponding to the original variable $\bar h$, (cf.~\eqref{vec_eom_h}), appears in the mass term.\footnote{We would like to emphasise that the ``mass" here is not quite what one understands under a mass of a scalar field in curved background. If this would be the case, the equations of motion should take the form $\Box\phi-m^2\phi=0$. The equations \eqref{veom_a} and \eqref{veoml} can be recast in such form by redefining the fields $a$ and $\lambda$. However, in this case the mass term becomes more complex and, in general, has two non-vanishing eigenvalues.} It was noted in~\cite{Blake:2013bqa}, that for time-independent configurations this implies the existence of a quantity that does not depend on $r$.

\begin{figure}[t]
\centering
\includegraphics[width=.4\linewidth]{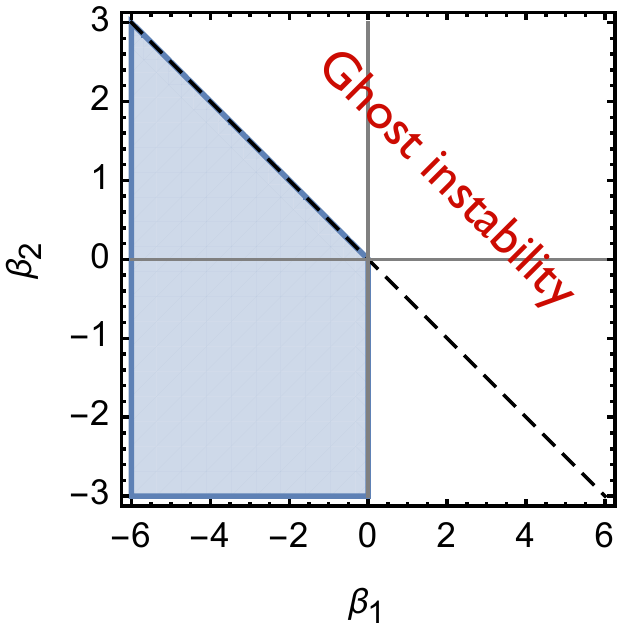}
\caption{Stability of the vector modes in the parameter space for the case $r_h = L$. The shaded region corresponds to the positive definite kinetic term of the vector modes and the absence of ghost instability. The dashed line corresponds to the ``wall of stability" found in~\cite{Vegh:2013sk}.}
\label{fig:par_plot}
\end{figure}

From the Lagrangian~\eqref{act_lam_a} one can see that the sign of the kinetic term for the field $\lambda$, descending from the gravity sector, is determined by the sign of the function $m^2(r)$. If $m^2(r)$ becomes negative anywhere in the region $0 \leq r \leq r_h$ the field $\lambda$ suffers from the ghost instability. Since $\lambda$ is coupled to the Maxwell field $a$, which has positive definite kinetic term, the instability shows up in the solutions of the equations of motion. This is exactly the instability found previously numerically in~\cite{Vegh:2013sk} and analytically in~\cite{Davison:2013jba} and~\cite{Blake:2013bqa}. The absence of ghost instability in the vector sector puts the following constraint on the parameters of the theory:
\be\label{cond2}
m^2(r)\equiv - \frac{\beta_1L}{r} - 2\beta_2 \leq 0 \quad \text{for } 0 \leq r \leq r_h \; .
\ee
It implies non-trivial conditions on the mass parameters $\beta_1$ and $\beta_2$:
\be\label{cond1}
\beta_1 \leq 0 \;, \quad \beta_2 \leq - \frac{L}{2 r_h}\beta_1 \; .
\ee
These constraints are presented in figure~\ref{fig:par_plot} for a particular choice of the black brane horizon location $r_h = L$ that is used in the conductivity calculations. The dashed line represents the ``wall of stability" found in~\cite{Vegh:2013sk}, which corresponds to the vanishing of $m^2(r)$ on the horizon.

\subsubsection{Equations of motion}
From the Lagrangian~\eqref{act_lam_a} one obtains two second order equations for $a$ and $\lambda$ 
\begin{align}\label{veom_a}
&(fa')'-\frac{\ddot a}{f}=\frac{r^2\mu^2}{r_h^2}a-\frac{2r^2\mu}{r_h}\lambda \; ,\\\label{veoml}
&\frac{1}{r^2}\left(\frac{r^2f}{m^2}\lambda '\right)'-\frac{\ddot\lambda}{fm^2} = \lambda - \frac{\mu}{2r_h}a \; ,
\end{align}
 with a coupled ``mass term'' with the mass matrix
\be\label{mmatrix}
\mathcal M=\begin{pmatrix}\frac{r^2\mu^2}{r_h^2} & -\frac{2r^2\mu}{r_h} \\ -\frac{\mu}{2r_h} & 1\end{pmatrix}.
\ee
This matrix has a vanishing determinant. In unitary gauge the equation \eqref{vec_eom_r} for the $i^{th}$ component of the vector fields becomes
\be
\dot\lambda_i= - \frac{m^2f}{2r^2}h_{ri} \; .
\ee
Upon inserting this in equations \eqref{veom_a} and \eqref{veoml} we recover the unitary gauge equations of motion as presented in~\cite{Blake:2013bqa}.

The original gauge invariant vector fields $\bar\pi_t,\,\bar\pi_r$, and $\bar h$ are then determined by the solutions for $a$ and $\lambda$ using the equations~\eqref{vec_eom_r}--\eqref{vec_eom_h}. These equations are derived in appendix~\ref{sec:vec_pert} and are given in equations \eqref{eqri}--\eqref{eqti}. We shall use these equations in order to compute the conductivity in the next section.

\section{Conductivity}\label{sec:conductivity}
Recently, massive gravity with two St\"uckelberg fields has been successfully used in order to describe the transport properties in materials with broken translational invariance via the AdS/CFT correspondence. By considering this type of massive gravity as the bulk gravitational theory on the black brane background, the calculated conductivity in the holographic framework is finite. This is different from the same calculation performed in the Einstein-Maxwell theory, where the diffeomorphism invariance in the bulk implies a translational symmetry in the field theory thus leading to infinite conductivity~\cite{Hartnoll:2009sz}. Instead, in massive gravity the background solutions for St\"ukelberg fields break the space-time diffeomorphisms, which allows for momentum dissipation in the boundary theory and renders the conductivity finite~\cite{Vegh:2013sk}. In this section we shall generalise these calculations to the diffeomorphism invariant case of massive gravity with two St\"uckelberg fields. From the boundary field theory point of view the St\"uckelberg fields induce new degrees of freedom, whose background values break the translational invariance (cf.~\cite{Blake:2013owa}).

The conductivity in AdS/CFT correspondence is calculated in the framework of the linear response theory. For a detailed reference on the calculation of the conductivity in the holographic framework see~\cite{Hartnoll:2009sz}. Here we shall present only the basic formulae needed for the calculations. At the background level of AdS/CFT, a boundary value of a bulk field gives rise to a background source for a corresponding dual field theory operator. In particular, the background bulk fields --- the metric $\hat g_{\mu\nu}$ and the Maxwell field $\hat A_\mu$ --- are sources for the energy-momentum tensor $T_{\mu\nu}$ and the current $J_\mu$ in the field theory. Considering small inhomogeneous perturbations around the background solutions allows us to find the retarded Green's function, which is defined to linearly relate the sources and the corresponding expectation values of the field theory operators. In the frequency space this relation can be stated as
\be
\delta\<\mathcal O_A\>(\omega, k)=G^R_{\mathcal O_A\mathcal O_B}(\omega,k)\delta \phi_B^{(0)}(\omega, k) \; ,
\ee
where we have collectively denoted the field theory operators as $\O_A=\{T_{\mu\nu},J_\mu,\dots\}$ and the boundary fields as $\phi_B^{(0)}=\{g_{\mu\nu}^{(0)},A_\mu^{(0)},\dots\}$ with the superscript $(0)$ indicating that the fields are evaluated at the AdS boundary, located at $r=0$. From the retarded Green's function of the transverse current $J_i$ we can determine the conductivity that relates the current density in a material to the electric field by the Kubo formula
\be\label{cond}
\sigma({\omega})=\frac{1}{i\omega}G^R_{J_iJ_i}(\omega) \; .
\ee
In terms of the Fourier space field components defined as
\begin{align}\label{fourier}
&\bar h_i(t,r)\equiv h(r)e^{i\omega t} \; , \\
&\bar\pi^i_r(t,r)\equiv-\frac{r^2}{f(r)}\pi(r)e^{i\omega t} \; , \\
&a_i(t,r)\equiv a(r)e^{i\omega t} \; ,
\end{align}
the retarded Green's function of the Maxwell field is given by 
\be
G^R_{J\,J}(\omega)=\left.\frac{a'}{a}\right|_{r=0}.
\ee
In the above expressions we have suppressed the transverse index $i$ for notational simplicity. The presented results are valid for both $x$ and $y$ directions independently. 

In order to find the response of the theory to small perturbations of the electrical field we need to find the solution of the equations of motion for the perturbations $a_i$. The field $a_i$ satisfies a set of coupled differential equations \eqref{eqri}--\eqref{eqti} given in the appendix. We see that the equation \eqref{eqri} is just a constraint equation relating the three fields and can be used in order to eliminate one of the fields.
In order to be able to compare to the results obtained previously in the unitary gauge, we would like to obtain a system of two coupled equations for the fields $\bar \pi^i_r$ and $a_i$. This is because in unitary gauge, when the St\"uckelberg field perturbations are set to zero, i.e. $\pi^i_{\textrm{unitary}}=0$, the gauge invariant fields are written as $\bar\pi^i_r=-h_{ri}^{\textrm{unitary}}$ and $a_i=a_i^{\textrm{unitary}}$. Hence, we can directly compare the resulting equations for the diffeomorphism invariant fields $\bar\pi^i$ and $a_i$ to the equations for the field perturbations $h_{ri}^{\textrm{unitary}}$ and $a_x^{\textrm{unitary}}$ in the unitary gauge, previously derived in~\cite{Vegh:2013sk}. This can be easily done after we substitute the ansatz \eqref{fourier}. After setting $L=r_h=1$ we obtain the following equations of motion
\begin{align}
\omega h-\mu r^2\omega a-\frac{i}{r}f(\beta_1+2r\beta_2)\pi &= 0 \; , \\ \label{feq1}
\frac{r^2}{\omega f}\left(-f\left[\omega\left(fa''+f'a'\right)-i\mu r(\beta_1+2r\beta_2)\pi\right]+\left(-\omega^3+\mu^2r^2\omega f\right)a\right) &= 0 \; , \\\label{feq2}
-r(\beta_1+2r\beta_2)\left[f\left(r(\beta_1+2r\beta_2)\pi'' + (\beta_1+4r\beta_2)\pi'\right) + r(\beta_1+2r\beta_2)f'\pi'\right] &-{} \nonumber\\
-\left[r^2\omega^2(\beta_1+2r\beta_2)^2-\beta_1^2f^2 + r(\beta_1+2r\beta_2)f\left((\beta_1+2r\beta_2)^2-\beta_1f'\right)\right]\pi &+{} \nonumber\\
+i\mu r^2\omega(\beta_1+2r\beta_2)^2fa &= 0 \; .
\end{align}
The last two equations can be solved numerically. For the regularity of the fields, we impose ingoing boundary conditions at the event horizon so that
\be\label{ing}
\pi(r), \; a(r) \, \sim \, e^{-i\omega/(4\pi T)\log(1-r)} \; ,
\ee
where $T$ is the temperature of the black hole defined in equation \eqref{temp}. From here we see that near the horizon the fields oscillate with an increasing frequency. This is, however, only due to the fact that we are working in a coordinate system in which the metric is singular at the horizon. For numerical stability we remove these near-horizon oscillations by choosing the following ansatz~\cite{Hartnoll:2009sz}: 
\be\label{bdy}
\pi(r) = f(r)^{-i\omega/(4\pi T)} P(r) \; , \qquad a(r) = f(r)^{-i\omega/(4\pi T)} A(r) \; .
\ee
In order to maintain the ingoing boundary conditions \eqref{ing} we require that in the vicinity of the horizon the functions $P(r)$ and $A(r)$ can be expanded in positive powers of $(1-r)$ as:
\begin{align}
\left.P(r)\right|_{r\to 1}=P_{\textrm{hor}}(r)\equiv p_0+p_1(r-1)+p_2(r-1)^2 + \dots \; , \\
\left.A(r)\right|_{r\to 1}=A_{\textrm{hor}}(r)\equiv a_0+a_1(r-1)+a_2(r-1)^2 + \dots \; .
\end{align}
The expansion coefficients $p_i$ and $a_i$ are determined by the near-boundary expansion of the equations of motion. In particular, $A_{\textrm{hor}}(1)=a_0, \ A_{\textrm{hor}}'(1)=a_1$, and $P_{\textrm{hor}}(1)=p_0 , \ P_{\textrm{hor}}'(1)=p_1$. Here we shall estimate only these first two expansion coefficients. For this one only needs to consider the zeroth order expansion of the equations \eqref{feq1}, \eqref{feq2} since they depend on $a_0, \, p_0, \, a_1, \, p_1$. By replacing $a_0\to A(1), \ a_1\to A'(1), \ p_0\to P(1), \ p_1\to P'(1)$ these equations can be used as ingoing boundary conditions. We also impose normalizability conditions at the boundary $A(0)=1, \ P(0)=0$. These are necessary because otherwise the field $P(r)$ diverges at the boundary. 

\begin{figure}
\centering
\begin{subfigure}{.495\textwidth}
  \centering
  \includegraphics[width=1\linewidth]{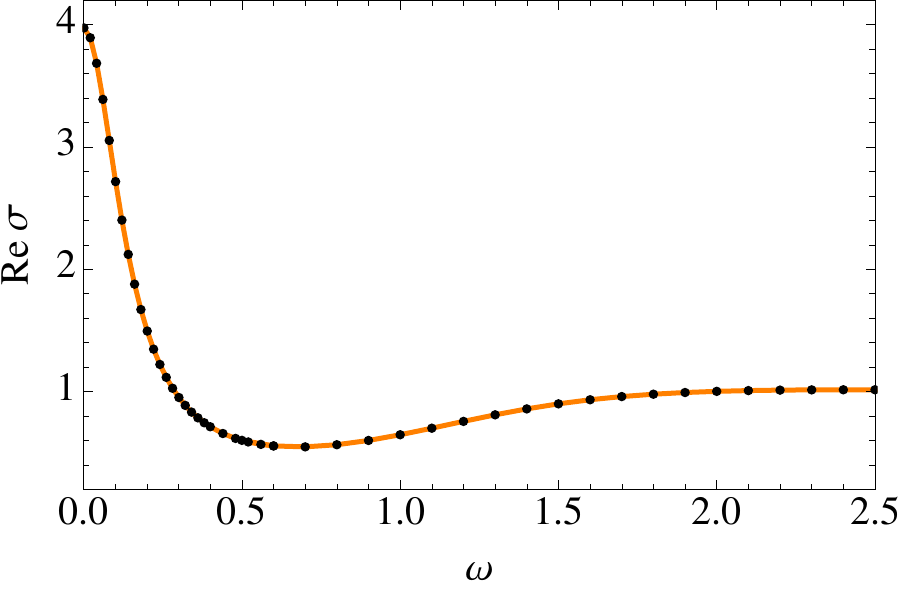}
  \caption{Real part}
  \label{fig:sub1}
\end{subfigure}
\begin{subfigure}{.495\textwidth}
  \centering
  \includegraphics[width=1\linewidth]{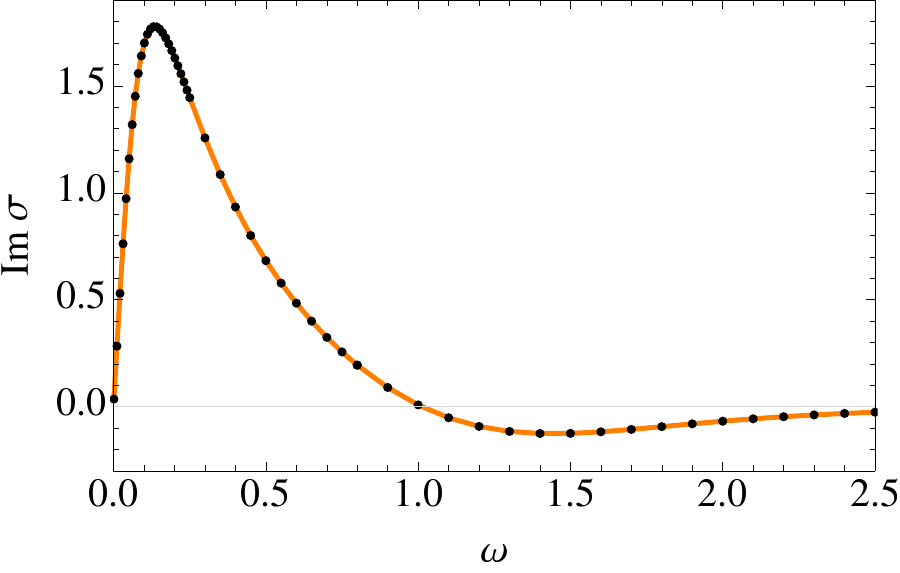}
  \caption{Imaginary part}
  \label{fig:sub2}
\end{subfigure}
\caption{The real and imaginary part of the electrical conductivity. The parameters are set to $\beta_1=-1,\ \beta_2=0, \ \mu=1.724$.}
\label{fig:cond}
\end{figure}

Finally, we numerically solve the equations \eqref{feq1} and \eqref{feq2} with the ansatz \eqref{bdy} and ingoing boundary conditions at the horizon $r=1$ and normalizable boundary conditions at the boundary  $r=0$ for the functions $A(r)$ and $P(r)$. The conductivity can then be found by evaluating the expression \eqref{cond} near the boundary
\be
\sigma(\omega)=\frac{1}{i\omega} \left.\frac{A'(r)}{A(r)}\right|_{r\to 0}-\frac{\beta_1}{8\pi T} \; .
\ee
The real and imaginary parts of the conductivity for a particular set of parameters are shown in figure \ref{fig:cond}. In order to be able to compare to previous results we have chosen the parameter values used in~\cite{Vegh:2013sk}. As we see from the figure, the real part of the conductivity exhibits the so-called Drude peak at zero frequency, which translates into the finiteness of the imaginary part of the conductivity. For comparison, in Einstein-Maxwell theory, the imaginary part of the conductivity has a pole at $\omega = 0$ and thus the real part of the conductivity is a delta function at $\omega=0$. This is a standard AdS/CFT result and is a well-known mismatch between the AdS/CFT predictions and the properties of real materials~\cite{Hartnoll:2009sz}. Hence, considering the reduced massive gravity as the gravitational theory in the bulk provides a framework of describing the momentum dissipation in the boundary field theory.

\section{Conclusions}\label{sec:conclusions}
In this work we have investigated the stability of the black brane solution in the reduced massive gravity with two St\"uckelberg fields. The interest in this question has arisen due to the recent idea that massive gravity with a degenerate reference metric can be used in order to incorporate momentum dissipation in holographic models~\cite{Vegh:2013sk}. The previous work on this topic has been done in the unitary gauge where the diffeomorphism invariance is broken in the direction of the transverse coordinates $\{x,y\}$. In this paper, we restore the diffeomorphism invariance in massive gravity by introducing two St\"uckelberg fields --- the Goldstone bosons of the broken coordinate transformations. 

Many aspects of the massive gravity with two St\"uckelberg fields have been investigated earlier in our work~\cite{Alberte:2013sma}. In particular, we have argued that this theory can be viewed as General Relativity with two minimally coupled scalar fields. As such, this type of massive gravity propagates four degrees of freedom --- two degrees of freedom of the helicity-two graviton and two degrees of freedom induced by the two scalars. By adding the gauge field, the total number of the dynamical degrees of freedom we expect raises to six --- the four of massive gravity and two of the Maxwell field. In this paper we have identified these degrees of freedom in a gauge invariant formulation and explored their stability. 

We have analysed the homogenous perturbations of the black brane solution in terms of the irreducible representations of the $SO(2)$ isometry group of the transverse coordinates. This allows us to split the field perturbations in scalar, vector, and tensor representations. In order to address the question of stability, we study the full quadratic action of the model. The scalar, vector, and tensor perturbations decouple from each other in the quadratic action and can be studied separately. We find two degrees of freedom in the tensor sector, four degrees of freedom in the vector sector, and none in the scalar sector. This gives six degrees of freedom in total as argued above. For inhomogeneous modes the decomposition in scalars, vectors, and tensors is different, and their presence can introduce additional conditions for the stability of the theory, cf.~\cite{Baggioli:2014roa}. We leave this for future research.

The two components of the dynamical tensor field correspond to the two degrees of freedom of the helicity-two graviton of the Einstein theory. Due to the dRGT-like mass term, the graviton has become massive with a space dependent mass $m^2_g(r)=-\beta_1r/2L$. Up to our knowledge, the tensor components of the graviton in the given model have not been studied elsewhere.

The story of the four degrees of freedom in the vector sector is more complex. The four dynamical fields are stored in two two-component vector fields. We obtain the quadratic Lagrangian~\eqref{act_lam_a} for the two propagating vector degrees of freedom.
The Lagrangian can be put in a form with diagonal derivative terms, but mixed ``mass matrix". This was already observed earlier in~\cite{Blake:2013owa} by studying the linearised equations of motion. The new result that we find is that the sign in front of the kinetic term of one of the two vector fields depends on a particular space dependent combination of the mass parameters $\beta_1$ and $\beta_2$. Concretely, the condition for the absence of negative energy modes, i.e. ghosts, reads
\begin{equation*}
m^2(r)=-2\beta_2-\frac{\beta_1L}{r}\geq 0 \; .
\end{equation*}
Since this should be satisfied for every $0 \leq r \leq r_h$, it implies a constraint on the mass parameters $\beta_1$ and $\beta_2$ of the form
\begin{equation*}
\beta_1 \leq 0 \;, \quad \beta_2 \leq - \frac{L}{2 r_h}\beta_1 \; .
\end{equation*}

Identical condition for the stability of the theory has been previously found numerically in~\cite{Vegh:2013sk} and analytically in~\cite{Davison:2013jba} and~\cite{Blake:2013bqa}. In the analytical studies it was found that the relaxation time of the momentum dissipation $\tau$ is inversely proportional to the combination $m^2(r_h)$. In the case of negative $m^2(r_h)$ this would result in the gain (instead of dissipation) of the momentum. Our findings reveal that the actual reason for this condition is that the theory propagates a ghost unless $m^2(r)\geq 0$. This is the main result of our paper. The observation that the condition $m^2(r) \geq 0$ has to be satisfied for all values of $r$ gives a new exclusion plot for the stability in the parameter space of $\beta_1$ and $\beta_2$, presented in figure~\ref{fig:par_plot}. 

Finally, we show that the equations of motion for the gauge invariant fields coincide with the equations of motion in the unitary gauge. We therefore claim that the previous results for the holographic conductivity in massive gravity obtained in unitary gauge in~\cite{Vegh:2013sk,Davison:2013jba,Blake:2013bqa} are actually diffeomorphism invariant. For completeness we have calculated numerically the optical conductivity to illustrate this fact. The results are shown in figure \ref{fig:cond}. As expected, the real and imaginary parts of the conductivity are finite and exhibit the so called Drude peak at low frequencies. 

\acknowledgments
We would like to thank Yegor Korovin for useful discussions on the subject. We also thank Richard Davison and Oriol Pujolas for valuable comments on the manuscript. This research was partially supported by the Israel Science foundation grant no. 239/10. The work of AK was also supported by the Kreitman foundation. The work of LA was also supported by the Minerva Foundation.

\appendix

 \section{Perturbative equations of motion}\label{sec:perturbations}
 In this section we find the first order perturbation equations from the equations of motion \eqref{eom}. The resulting equations are equivalent to the ones derived from the action in section \ref{sec:action} and, therefore, serve only as a cross check of the obtained results. We separate the different parts of the equation of motion \eqref{eom} as 
 \be
 G_{\mu\nu}-g_{\mu\nu}\Lambda+X_{\mu\nu}=T_{\mu\nu} \; ,
 \ee
 where $X_{\mu\nu}$ is defined in equation \eqref{xmunu} and
 \begin{align}
 &G_{\mu\nu} = R_{\mu\nu}-\frac{1}{2}g_{\mu\nu}R \; ,\\
 &T_{\mu\nu}=-\frac{L^2}{2}\left(F_{\mu\alpha}F^\alpha_{\ \nu}+\frac{1}{4}g_{\mu\nu}F_{\alpha\beta}F^{\alpha\beta}\right) .
 \end{align}
  Each part can be perturbed separately. The variations of the Einstein tensor and of the energy-momentum tensor of the Maxwell field are well known. Below we present the main steps for the variations of the graviton mass term. 
 
\subsection{Variation of the graviton mass term}
In order to find the variations of the part of the equations of motion arising from the graviton mass term $X_{\mu\nu}$ we shall first determine the perturbations of the square root matrix $\sqrt\I^A_B$. The latter is defined so that it satisfies 
\be
\sqrt\I^A_C\sqrt\I^C_B=\I^A_B=g^{\mu\lambda}\pt_\mu\phi^A\pt_\nu\phi^C\delta_{BC} \; .
\ee
Let us perturb the matrices $\sqrt\I^A_B$ and $\I^A_B$ so that the above relation becomes
\begin{align}\label{expansion}
\left(^{(0)}\sqrt\I^A_C+\,\,^{(1)}\sqrt\I^A_C+\dots\right)&\left(^{(0)}\sqrt\I^C_B+\,^{(1)}\sqrt\I^C_B+\dots\right)=^{(0)}\I^A_B+\,^{(1)}\I^A_B+\dots \; .
\end{align}
By expanding both sides of the equation \eqref{expansion} and comparing the first order terms we find that the first order perturbations of the square root matrix are determined by the equation
\be\label{pert}
\delta\sqrt\I^A_C\,\sqrt{\hat\I}^C_B+\sqrt{\hat\I}^A_C\,\delta\sqrt\I^C_B=\delta \I^A_B \; .
\ee
Since the background matrix $\hat\I^A_B$ is diagonal, the above relation simplifies to
\be\label{pertI2}
\delta\sqrt\I^A_B=\frac{L}{2r}\delta\I^A_B \; .
\ee 

The variations of $X_{\mu\nu}$ defined in \eqref{xmunu} can then be found by using the definitions of $\U_{1,2}$ and $\mathcal E_{\mu\nu},\,\mathcal E^2_{\mu\nu}$ given in equations \eqref{defy}, \eqref{defiab} and \eqref{defe1}--\eqref{defe2} together with the variations of $\delta\I^A_B$ and $\delta\sqrt\I^A_B$ given in \eqref{pertI1}, \eqref{pertI2}. Below we give the final expressions for the complete first order equations of motion for the vector perturbations in terms of the diffeomorphism invariant variables defined in section \ref{sec:vectors}. 

\subsection{Vector perturbations}\label{sec:vec_pert}
The equations for vector perturbations consist from the equation of motion for the Maxwell field, the equation of motion for the St\"uckelberg fields, and the $(ti)$ and $(ri)$ components of the perturbed Einstein equations. 

The equations of motion for the St\"uckelberg fields $\phi^A$ we obtain by varying their action 
\be
S_{\phi} = \frac{\M^2m^2}{2} \int \sqrt{-g} \Big(\beta_1 \U_1(\sqrt{\I}) + \beta_2\, \U_2(\sqrt{I})\Big) \; ,
\ee
By using the explicit expressions for the characteristic polynomials of the square root matrix given in \eqref{u1u2} we find the following equation of motion for the scalar fields:
\be
\frac{\delta S_\phi}{\delta \phi^C}=-\pt_\mu\left[\frac{\beta_1}{\U_1}\sqrt{-g}g^{\mu\nu}\pt_\nu\phi_C+2\sqrt{-g}g^{\mu\nu}\pt_\nu\phi_B\varepsilon_{CA}\varepsilon^{BD}\I^A_D\left(\frac{\beta_1}{\U_1\U_2}+\frac{2\beta_1}{\U_2}\right)\right] \; .
\ee
By substituting $\phi^A=x^\mu\delta^A_\mu +\pi^A$ we obtain an equation for the scalar field perturbations $\pi^A$. After combining this equation with the equations for the Maxwell field and metric perturbations and rewriting them in terms of the gauge invariant fields $\bar h_i$, $a_i$, and $\bar\pi^i_r$ defined in \eqref{vec}, we arrive at the following equations of motion:
\begin{align}\label{eqri}
(ri):\quad &\left(2\beta_2+\frac{\beta_1L}{r}\right)\bar\pi^i_r+\frac{\mu r^2}{r_hf}\dot{ a}_i-f^{-1}\,\dot{\bar h}_i = 0 \; ;\\
\label{eqa}
(a_i):\quad &\ddot{ a}_i-f^2 a_i''-ff'a_i'+\frac{\mu f}{r_h}\bar h_i = 0 \; ;\\
\label{eqpi}
(\bar\pi^i_r):\quad &-\ddot{\bar\pi}^i_r+\dot{\bar h}_i+f^2\bar\pi^i_r\,''+\left(3ff'-\frac{(3\beta_1L+4\beta_2r)f^2}{r(\beta_1L+2\beta_2r)}\right)\bar\pi^i_r\,' + {} \nonumber\\
&+\left(f'^2-\frac{-\mu^2r^3+(\beta_1L+2\beta_2r)r_h^2}{rr_h^2}f\right)\bar\pi^i_r - {} \nonumber\\
&-\frac{4(\beta_1L+\beta_2r)ff'}{r(\beta_1L+2\beta_2r)}\bar\pi^i_r+\frac{(3\beta_1^2L^2+12\beta_1\beta_2Lr+8\beta_2^2r^2)f^2}{r^2(\beta_1L+2\beta_2r)^2}\bar\pi^i_r=0 \,; \\
\label{eqti}
(ti),\,(ri):\quad &\textrm{equivalent to }(\bar\pi^i_r)\textrm{ equation} \; .
\end{align}
We see that there are two independent second order equations. Since each of the fields has two components, then these equations describe four propagating degrees of freedom. Two of them are the degrees of freedom of the Maxwell field and the other two are gravitational degrees of freedom induced by the two St\"uckelberg fields.

\section{Comparison to the \texorpdfstring{$\sqrt{g^{-1}f}^\mu_\nu$}{} formalism}\label{sec:app1}
In the previous works on massive holography another formalism of writing the graviton mass term has been used (see~\cite{Vegh:2013sk,Blake:2013bqa,Davison:2013jba}). In particular, the potential for the graviton in action \eqref{action} was written as
\be
\U(g,\phi)=\sum_{i=1}^{4}\beta_i \, \U_i(\Omega) \; ,
\ee
where as before $\U_i(\mathbb X)$ are the characteristic polynomials of some matrix $\mathbb X^\mu_\nu$ defined as $\U_1= \[\mathbb X\],\,\U_2=\[\mathbb X\]^2-\[\mathbb X^2\],$ and
\begin{align}
\U_3 &= \[\mathbb X\]^3-3\[\mathbb X\]\[\mathbb X^2\]+2\[\mathbb X^3\] \; , \\
\U_4 &= \[\mathbb X\]^4-6\[\mathbb X^2\]\[\mathbb X\]^2+8\[\mathbb X^3\]\[\mathbb X\]+3\[\mathbb X^2\]^2-6\[\mathbb X^4\] \; ,
\end{align}
where $\[\mathbb X\]\equiv\mathbb X^\mu_\mu,\,\[\mathbb X^2\]\equiv\mathbb X^\mu_\nu\mathbb X^\nu_\mu,\dots$. In distinction from section \ref{sec:model}, the matrix $\Omega$ is defined as $\Omega^\mu_\nu\equiv\sqrt{g^{\mu\alpha}f_{\alpha\nu}}\equiv\sqrt{g^{-1}f}^\mu_\nu$. \footnote{The field $\Omega^\mu_\nu$ equals to the $\K^\mu_\nu$ of~\cite{Vegh:2013sk}. } This is a $4\times 4$ matrix with the reference metric $f_{\mu\nu}$ parametrized by the St\"uckelberg fields as $f_{\mu\nu}=\partial_\mu\phi^A\partial_\nu\phi^Bf_{AB}(\phi)$ where $f_{AB}$ is the reference metric in the configuration space of scalar fields. 

In~\cite{Vegh:2013sk} a degenerate reference metric was considered such that in the unitary gauge it takes the form 
\be\label{funit}
f_{\mu\nu}=\mathrm{diag}(0,0,c^{2},c^{2}) \; ,
\ee
where $c$ is a constant and the spacetime index $\mu=\{t,r,x,y\}$. Rewritten in a diffeomorphism invariant form this is equivalent to working with only two St\"uckelberg fields $\phi^A$, $A=\{2,3\}$ with a reference metric $f_{AB}=\delta_{AB}=\mathrm{diag}(1,1)$ as was done in this paper. In the case of a degenerate reference metric \eqref{funit}, the matrix $\Omega^\mu_\nu$ has only two non-zero eigenvalues and therefore the characteristic polynomials $\U_3(\Omega),\,\U_4(\Omega)$ vanish, and the graviton mass term reduces to
\be
\U(g,\phi)=\beta_1 \, \U_1+\beta_2 \, \U_2 \; .
\ee
The reference metric $f_{\mu\nu}$ takes the diagonal form \eqref{funit} only on the background solution 
\be\label{solsc}
\hat\phi^A\equiv c\,x^\mu\delta^A_\mu \; .
\ee
If one considers perturbations of the St\"uckelberg fields defined as $\pi^A\equiv \phi^A-\hat\phi^A$, the field $f_{\mu\nu}$ is different from \eqref{funit}. By performing an infinitesimal coordinate transformation the scalar field perturbations can be set to zero. This is the unitary gauge. By working in the unitary gauge we fix two out of four coordinate transformations and therefore the resulting theory has a partially broken diffeomorphism invariance. 

The background solution of the metric is determined by the equation of motion~\eqref{eom} with $X_{\mu\nu}$ defined as
\be\label{mass}
X_{\mu\nu}=-\frac{\beta_1}{2}\left(g_{\alpha\beta}\,\U_1-\Omega_{\alpha\beta}\right)-\beta_2\left(\Omega^2_{\alpha\beta}-\[\Omega\]\Omega_{\alpha\beta}+\frac{1}{2}g_{\alpha\beta}\,\U_2\right) \; .
\ee
To derive this expression we have used the following relations and notations:
\begin{align}
&\Omega_{\alpha\beta}\equiv\Omega^\gamma_{(\beta} g_{\gamma\alpha)} \; , \quad 
\Omega^2_{\alpha\beta}\equiv\Omega^\gamma_\nu\Omega^\nu_{(\beta}g_{\gamma\alpha)} \; , \quad 
\frac{\delta\[\Omega^n\]}{\delta g^{\alpha\beta}} = \frac{n}{2}\Omega^n_{\alpha\beta} \; .
\end{align}
These equations of motion for the background metric are equivalent to the ones obtained in section \ref{sec:background} and, hence, they admit the same AdS black brane solution given in equations \eqref{solmetric} and \eqref{fr}.

\subsection{Perturbations of the graviton mass term}
In order to find the first order perturbation of the equations of motion we consider the parts due to the Einstein gravity, Maxwell term, and graviton mass separately. The variation of the Einstein tensor and the Maxwell energy-momentum tensor is standard. The only non-trivial step is the variation of the terms arising from the graviton mass. For completeness we present the main steps of the derivation below. 

We shall first find the perturbations of the square root matrix $\Omega^\mu_\nu$ defined as 
\be
\Omega^\mu_\alpha \, \Omega^\alpha_\nu = g^{\mu\lambda}f_{\lambda\nu} \; .
\ee
By perturbing the matrices $\Omega^\mu_\nu$, $g^{\mu\nu}$ and $f_{\mu\nu}$ the above relation becomes
\begin{align*}
\left(^{(0)}\Omega^\mu_\alpha+\,^{(1)}\Omega^\mu_\alpha+\dots\right)&\left(^{(0)}\Omega^\alpha_\nu+\,^{(1)}\Omega^\alpha_\nu+\dots\right)=g^{\mu\lambda}f_{\lambda\nu}\\
=&\left(^{(0)}g^{\mu\lambda}+\,^{(1)}g^{\mu\lambda}+\dots\right)\left(^{(0)}f_{\lambda\nu}+\,^{(1)}f_{\lambda\nu}+\dots\right).
\end{align*}
After expanding both sides of the equation and comparing the terms of the same order in perturbations one finds recursive relations which allow one to find the perturbative expansion of the square root matrix order by order. Since here we are interested only in the perturbations up to linear order, then we denote $\delta\Omega^\mu_\nu\equiv\,^{(1)}\Omega^\mu_\nu$, $\delta g^{\mu\nu}\equiv\,^{(1)} g^{\mu\nu}$, and $\delta f_{\mu\nu}\equiv\,^{(1)} f_{\mu\nu}$. The perturbations of the square root matrix are then determined by the equation
\be\label{pert_om}
\delta\Omega^\mu_\alpha\,\hat\Omega^\alpha_\nu+\hat\Omega^\mu_\alpha\,\delta\Omega^\alpha_\nu=\delta g^{\mu\alpha}\,\hat f_{\alpha\nu}+\hat g^{\mu\alpha}\,\delta f_{\mu\nu} \; .
\ee
Around the background solution \eqref{solsc} of the St\"uckelberg scalar fields $\phi^A$ the reference metric $f_{\mu\nu}=\partial_\mu\phi^A\partial_\nu\phi^B\delta_{AB}$ can be expanded as
\begin{align*}
f_{\mu\nu} &=\hat f_{\mu\nu}+\delta f_{\mu\nu}+\mathcal O(\pi^2) =c^2\delta_\mu^A\delta_\nu^B\delta_{AB}+c\left(\delta_\mu^A\pt_\nu\pi^B+\pt_\mu\pi^A\delta_\nu^B\right)\delta_{AB}+\mathcal O(\pi^2) \; .
\end{align*}
Hence, on the background \eqref{solmetric} the square root matrix $\Omega^\mu_\nu$ equals to $\hat\Omega^\mu_\nu=(cr/L)\delta^\mu_A\delta^A_\nu$.
 By defining the metric perturbations as $g_{\mu\nu}=\hat g_{\mu\nu}+\delta g_{\mu\nu}$ and the corresponding first order perturbations of the inverse metric as $\delta g^{\mu\nu}=-\hat g^{\alpha\mu}\hat g^{\beta\nu}\delta g_{\alpha\beta}$, we arrive at the expression for the first order perturbations of the square root matrix
 \be\label{pertsqrt}
  \delta \Omega^\mu_\nu=\begin{pmatrix}
  \delta\Omega^t_t & \delta\Omega^t_r & \frac{cL}{r}\delta g^{tx}& \frac{cL}{r}\delta g^{ty} \\
  & & &\\
  \delta\Omega^r_t & \delta\Omega^r_r & \frac{cL}{r}\delta g^{rx} & \frac{cL}{r}\delta g^{ry} \\
    & & &\\
  0  & 0  & \frac{cL}{2r}\delta g^{xx}+\frac{r}{L}\partial_x\pi^x & \frac{cL}{2r}\delta g^{xy}+\frac{r}{2L}(\partial_y\pi^x+\partial_x\pi^y)  \\
  & & &\\
  0 & 0 & \frac{cL}{2r}\delta g^{xy}+\frac{r}{2L}(\partial_y\pi^x+\partial_x\pi^y) & \frac{cL}{2r}\delta g^{yy}+\frac{r}{L}\partial_y\pi^y
 \end{pmatrix}.
 \ee
 Here the upper index denotes the row, and the down index denotes the column of the matrix $\delta\Omega^\mu_\nu$. We note, that the components in the upper left $2\times 2$ matrix are not determined by the equations \eqref{pert_om} due to the degeneracy of the background values of $\hat\Omega^\mu_\nu$ and $\hat f_{\mu\nu}$. It is therefore why the variables $\mathcal I^{AB}=g^{\mu\nu}\partial_\mu\phi^A\partial_\nu\phi^B$ seem to be a better choice for writing the Lagrangian for the graviton mass term. The equations of motion in the unitary gauge however are the same in both cases. 
 
 With \eqref{pertsqrt} at hand it is then straightforward to find the variation of the equations of motion arising from the graviton mass term $X_{\mu\nu}$. 
The only subtlety is to symmetrize the variations $\delta\Omega_{\mu\nu}$ and $\delta\Omega^2_{\mu\nu}$ in $\mu$ and $\nu$.  

\subsection{Final equations}
We use the ansatz, where only the sector of vector perturbations in one of the transverse directions is considered
\be
\delta g_{\mu\nu}=\begin{pmatrix}
  0 & 0 & \delta g_{tx}(r)&0 \\
  0 & 0 & \delta g_{rx}(r) & 0 \\
  \delta g_{tx}(r)  & \delta g_{rx}(r)  & 0 & 0  \\
  0 & 0 & 0 & 0
 \end{pmatrix} e^{i\omega t} \; , \quad a_\mu=\left(0,0,a_x(r)e^{i\omega t},0\right) \; ,
 \ee
 where $a_\mu\equiv A_\mu-\hat A_\mu$ is the perturbation of the Maxwell field. The only non-zero contribution due to the variations of the Einstein tensor arise in the following components of the Einstein equations
\begin{align*}
(tx):\quad &-\delta g_{tx}\left[-2r^2r_h^2f''+8rr^2_hf'-16r_h^2f+4\beta_1cLrr_h^2+4\beta_2c^2r^2r_h^2 + {} \right. \nonumber \\
&+\left.12r_h^2+\mu^2r^4\right]+2rr_hf\[{L^2\mu ra_x'}+r_h\left(ir\omega\delta g_{rx}'-2\delta g_{tx}'-r\delta g_{tx}''\right)\] = 0 \; ;\\
(rx):\quad&\delta g_{rx}fr_h^2\[-2r^2f''+8rf'-12f+4\beta_1cLr+4\beta_2c^2r^2+\frac{\mu^2r^4}{r_h^2}+12\] + {} \nonumber \\
&+2\delta g_{rx}r^2r_h^2\omega^2+2irr_h^2\omega\left(2\delta g_{tx}+r\delta g_{tx}'\right)-{2iL^2\mu r^2r_h\omega a_x} = 0 \; .
\end{align*}
The non-zero contributions to the components $(tt),\,(rr),\,(tr),\,(xx),\,(yy)$ are entirely do to the variation of $X_{\mu\nu}$. These allow one to set the undetermined components $\delta \Omega^t_t,$ $\delta\Omega^r_r,$ $\delta \Omega^r_t,\,\delta\Omega ^t_r$ to zero. The above equations coincide with the equations presented in~\cite{Vegh:2013sk} (with $F=m^2=1$ and $\alpha=\beta_1,\,\beta=\beta_2$) after we rescale the Maxwell field perturbations as $a_x\to 2 a_x$. This difference was also noted in~\cite{Davison:2013jba}. Combining these equations with the equation of motion for the Maxwell field \eqref{eqa} allows one to eliminate $\delta g_{tx}$. Then, upon replacement $\delta g_{rx}=\delta\tilde g_{rx}/f(r)$ and identification $\delta\tilde g_{rx}=\pi$ and $a_x=a$, the resulting equations are equivalent to equations \eqref{feq1} and \eqref{feq2}.

\end{document}